\documentclass[12pt,preprint,iop]{emulateapj}

\usepackage{color}
\usepackage{graphicx}
\usepackage{fancyheadings}

\shorttitle{Searching for Cooling Signatures in Strong Lensing Galaxy Clusters}
\shortauthors{Blanchard et al.}

\begin{document}

%%%%%%%%%%%% TITLE  %%%%%%%%%%%%%%%%%%%%%%%%%%%%%%%%%%%%%%%%%%%%%%%%% 
% page: title

\title{Searching for Cooling Signatures in Strong Lensing Galaxy Clusters: Evidence Against Baryons Shaping the Matter Distribution in Cluster Cores}

\author{Peter K. Blanchard\altaffilmark{1}$^{,}$\altaffilmark{2}, Matthew B. Bayliss\altaffilmark{2}$^{,}$\altaffilmark{3}, Michael McDonald\altaffilmark{4}$^{,}$\altaffilmark{5}, H{\aa}kon Dahle\altaffilmark{6}, Michael D. Gladders \altaffilmark{7}$^{,}$\altaffilmark{8}, Keren Sharon\altaffilmark{9}, Richard Mushotzky\altaffilmark{10}}
\email{pblanchard@berkeley.edu}

\altaffiltext{1}{University of California, Berkeley, Astronomy
  Department, B-20 Hearst Field Annex 3411, Berkeley, CA 94720-3411} 
\altaffiltext{2}{Harvard-Smithsonian Center for Astrophysics, 60
  Garden St., Cambridge, MA 02138}
\altaffiltext{3}{Harvard University, Department of Physics, 17 Oxford
  St., Cambridge, MA 02138}
\altaffiltext{4}{Massachusetts Institute of Technology, Kavli
  Institute for Astrophysics and Space Research, 77 Massachusetts
  Ave. 37-287, Cambridge, MA 02139}
\altaffiltext{5}{Hubble Fellow}
\altaffiltext{6}{Institute of Theoretical Astrophysics, University of Oslo,
P.O. Box 1029, Blindern, N-0315 Oslo, Norway}
\altaffiltext{7}{Department of Astronomy \& Astrophysics,
University of Chicago, 5640 South Ellis Avenue, Chicago, IL 60637}
\altaffiltext{8}{Kavli Institute for Cosmological Physics,
University of Chicago, 933 East 56th Street, Chicago, IL 60637}
\altaffiltext{9}{Department of Astronomy, University of Michigan,
500 Church Street Ann Arbor, MI 48109-1042}
\altaffiltext{10}{Astronomy Department, University of Maryland, College Park, MD 20742, USA}
%%%%%%%%%% ABSTRACT %%%%%%%%%%%%%%%%%%%%%%%%%%%%%%%%%%%%%%%%%%%%%%%%%%% 
% page: abstract

\begin{abstract}
\medskip
The process by which the mass density profile of certain galaxy
clusters becomes centrally concentrated enough to produce high strong lensing
(SL) cross-sections is
not well understood.  It has been suggested that the baryonic
condensation of the intra-cluster medium (ICM)
due to cooling may drag dark matter to the cores and thus steepen the
profile.  In this work, we search for evidence of ongoing ICM cooling in
the first large, well-defined sample of strong lensing selected
galaxy clusters in
the range $0.1 < z < 0.6$. Based on known correlations between the ICM
cooling rate and both optical emission line luminosity and star formation, we
measure, for a sample of 89 strong lensing 
clusters, the fraction of clusters that have [OII]$\lambda\lambda$3727 emission in their brightest 
cluster galaxy (BCG).  
We find that the fraction of line-emitting BCGs is constant as a function of redshift for $z > 0.2$ 
and shows no statistically significant deviation from the total cluster
population.  Specific star formation rates, as traced by the strength
of the 4000\AA\ break, D$_{4000}$, are also consistent with the general cluster 
population. Finally, we use optical imaging of the SL clusters to measure the 
angular separation, R$_{arc}$, between the arc and the center of mass of each lensing 
cluster in our sample and test for evidence of changing 
[OII] emission and D$_{4000}$ as a function of R$_{arc}$, a proxy observable for SL cross-sections. 
D$_{4000}$ is constant 
with all values of R$_{arc}$, and the [OII] emission fractions show no 
dependence on R$_{arc}$ for R$_{arc} > 10^{\prime\prime}$ and only very marginal 
evidence of increased weak [OII] emission for systems with R$_{arc} < 10^{\prime\prime}$. 
These results argue against the ability of baryonic cooling associated with cool core activity
in the cores of galaxy clusters to strongly modify the underlying 
dark matter potential, leading to an increase in strong lensing
cross-sections.     
\smallskip
\end{abstract}


\section{Introduction}
Galaxy clusters that exhibit
strong lensing in their cores are some of the rarest
objects in the Universe and the global strong lensing
cross-section for galaxy cluster-scale structures is dominated by a 
small fraction of the total galaxy cluster population.  In strong lensing (SL) 
galaxy clusters, theory and simulations predict that certain astrophysical 
factors play a role in increasing SL cross-sections. $N$-body 
simulations predict that dark matter concentrations in strong 
lensing clusters should be significantly larger than most other clusters 
\citep{Hennawi2007,  Meneghetti2010} and that triaxiality and 
clumpiness in the cores could be significant in producing SL clusters
\citep{Hennawi2007}.  While many strong lensing clusters have high
mass, \citet{Dalal2004} showed that the central mass concentration
rather than the mass itself is a more important determinant of how 
giant arcs are produced by cluster-scale mass distributions. 
However, many studies have found that simple dissipationless (i.e. dark matter only) 
cosmological simulations tend to under-predict the abundance of SL 
galaxy clusters by an order of magnitude or more indicating that all factors such as triaxiality and substructure contributing to large strong lensing cross-sections have not been taken into account 
\citep[e.g.][]{Bartelmann1998,luppino1999,ZaritskyGonzalez2003,Gladders2003,li2006}.

Additional factors that may contribute to large cross-sections
include dark matter condensation due to cooling baryons
\citep{Rozo2008, Mead2010}, central
galaxies and substructure \citep{Flores2000, Meneghetti2000,
  Meneghetti2003, Hennawi2007, Meneghetti2010}, triaxiality of cluster
mass profiles \citep{Oguri2003, Dalal2004, Hennawi2007, Meneghetti2010},
major mergers that increase the cross-section on short timescales
\citep{Torri2004, Fedeli2006, Hennawi2007},
structure along the line of sight not related to the lens or source
\citep{Wambsganss2005, Hilbert2007, PuchweinHilbert2009}, and the
properties of the background galaxies \citep{HamanaFutamase1997,
  Wambsganss2004, Bayliss2011, Bayliss2012}.  \citet{Wambsganss2004} and \citet{Dalal2004} showed that
increasing the source redshifts in simulations increases SL
cross-sections.  Failure to account
for realistic source redshift distributions has been demonstrated to have 
a factor of $\sim10\times$ effect on giant arc abundances \citep{Bayliss2012}.  Using SL clusters to test predictions 
from theories and cosmological models 
has historically been limited by the lack of large, well-defined lens 
samples. The first homogeneously selected cluster lens samples had 
sizes N$\backsim$5 \citep{LeFevre1994, ZaritskyGonzalez2003, Gladders2003} 
and thus too small to have statistical power, but this is now changing as we 
move solidly into a new era of wide-field 
imaging surveys -- such as the SDSS 
\citep[e.g.,][]{Hennawi2008,kubo2009,diehl2009,Kubo2010,Bayliss2011S,Oguri2012}, 
the Canada-France-Hawaii-Telescope Legacy Survey \citep[CFHTLS;][]{cabanac2007}, 
and the Second Red Sequence Cluster Survey \citep[RCS2;][]{Bayliss2012}.

One reasonable physical scenario that could contribute to the accumulation 
of mass in the cores of strong lensing galaxy clusters involves
baryonic cooling.  The hot intracluster medium (ICM) in clusters cools by losing energy in the form of X-ray radiation.  In this picture, in order to maintain hydrostatic
equilibrium, the cool gas flows inward establishing a cooling flow \citep[e.g.][]{Fabian1994}. 
In some galaxy clusters, the cooling rate in the center is anomalously
high to the point that the cooling time is shorter than the Hubble
time.  Classical estimates suggest cooling rates of about 1000
M$_{\Sun}$/yr which should lead to equally high star formation rates.
However, such dramatic amounts of star formation are not observed so
there must be some mechanism, such as feedback from active galactic nuclei,
which can offset the energy loss from cooling \citep{McNamara2007, Fabian2012, McNamara2012}.  
Even so, there is often a small amount of cooling gas fueling star formation in these ``cool core" clusters, representing the residual in the feedback/cooling balance, at typical levels of 1-10 M$_{\Sun}$/yr \citep{O'Dea2008, McDonald2011b}, but can be as high as $>$100 M$_{\Sun}$/yr \citep{McNamara2006, O'Dea2008, McDonald2012pheonix}.

Several studies have found that cool core clusters also contain
optical emission-line nebulae in the central regions \citep{Hu1985,
Johnstone1987, Heckman1989, Edwards2007, Hatch2007, McDonald2010,
McDonald2011a}.  In addition, star formation rates (SFR)
in the brightest cluster galaxies (BCGs) of cool core clusters are known to be higher than SFRs in
non-cool cores \citep{Johnstone1987, McNamaraOConnell1989, Allen1995} and excess IR emission has been found to be proportional to H$\alpha$ emission
suggesting both may be due to star formation as a result of the
cooling intracluster medium \citep{O'Dea2008}.  By studying UV and H$\alpha$ emission of extended filaments in cool
cores, \citet{McDonald2011b, McDonald2012} found that in the majority of clusters (with Perseus as a notable exception), the warm gas is primarily photoionized
by massive, young stars, with small contributions most likely from slow shocks.  Both the optical emission and the star formation seem to be related
to the X-ray properties of the ICM, such as the X-ray cooling rate, suggesting
that cooling gas from the intracluster medium is the source of the
warm ionized gas and the fuel for star formation \citep[e.g.,][]{Edge2001, O'Dea2008, McDonald2010, McDonald2011a, McDonald2011b, Tremblay2012}.  This link suggests that the presence of either warm ionized gas or ongoing star formation in the BCG may indicate that the ICM is cooling rapidly in the cluster core.  Recent work demonstrates that the evolution
of cool core clusters matches the evolution of optically
emitting nebulae, suggesting that optical emission-line nebulae may
serve as an effective tracer for cool cores \citep{Donahue1992, McDonald2011,
  Samuele2011}.  This is significant because at high redshifts it is
difficult to determine the cooling rate since the X-ray flux is very low for most of the sources. 
This paper makes use of the correlation
between optical emission-line luminosity and cool core strength, the former having the advantage of
being measureable from the ground via even modest aperture telescopes.

In this work we use observations of a sample of 89 strong lensing 
galaxy clusters with BCG spectra available from the SDSS to test for evidence 
that baryonic cooling is contributing strongly to the high surface
mass density of strong lensing galaxy clusters. This paper is organized as 
follows.  In Section 2 we describe the strong lensing 
cluster sample and the data analyzed.  In Section 3, we describe
our analysis methods and present the evolution of [OII] line
emission and 4000\AA\ break ratio for our sample compared to the total
cluster population. Section 4 provides a discussion of the results
and the paper concludes with a summary in Section 5.

In this paper we assume $\Omega_{M} = 0.27$, $\Omega_{\Lambda} =
0.73$, and H$_{0} = 71$ km s$^{-1}$ Mpc$^{-1}$ \citep{Hinshaw2009}.

\section{Cluster Sample and SDSS Data}

\subsection{Strong Lensing Selected Cluster Sample}
To minimize systematic effects and to allow statistically robust analysis it is important that we have a large, uniformly selected sample of
SL clusters. In an attempt to obtain a well-understood sample, a
systematic search for strong lensing galaxy clusters in the SDSS DR7
was carried out \citep{Hennawi2008}. Follow-up observations and analyses of 
subsets of the Sloan Giant Arcs Survey (SGAS) sample have been previously published \citep{Bayliss2010, Koester2010, Bayliss2011S, Bayliss2011, Oguri2012}.  In brief, 
candidate galaxy clusters in the SDSS data were selected at optical wavelengths using 
the red sequence algorithm \citep{Gladders2000}.  Each candidate optically 
selected galaxy cluster was visually inspected by
four experts who each assigned a numerical score based on the presence or 
absence of any evidence of strong lensing in the images. The score scale ranges 
from 0 to 3 where 3 means there is obvious strong lensing and 0 means no
evidence for lensing.  The final score is the average of
each individual score from each person.  Follow-up observations
were obtained so that the purity of the sample, the number of candidate strong lenses
that actually are SL clusters, could be understood. 
These efforts have produced the first sample of hundreds of candidate strong
lensing galaxy clusters, which will be described in full detail in a forthcoming
publication (M. D. Gladders et al., in preparation).

We are using this new large sample of SL clusters to conduct the
first systematic search for observational evidence of enhanced gas cooling in
strong lensing galaxy clusters.  The completeness and purity as a
function of score for this SL sample is well understood and the
majority of the sample clusters have deep, optical follow-up observations 
(98\% follow-up for score $>$ 1.5 and 75\% follow-up for score $>$ 1.0).  
We remove from the sample those clusters for which the score is below 1.3 
to prevent clusters that do not clearly exhibit strong lensing from contaminating our
conclusions because the purity of the sample as a function of score
drops off strongly between mean scores of 1.5 and 1.0.
After a visual inspection of deep follow-up images of each cluster, we
also removed 6 
SL cluster candidates that cannot be visually confirmed at high 
confidence as lenses.  As described in detail in the next section, we then 
match the remaining clusters to the SDSS 
spectroscopic catalog using updated coordinates from follow-up data.

\begin{figure}[ht]
\epsscale{1.1}
\plotone{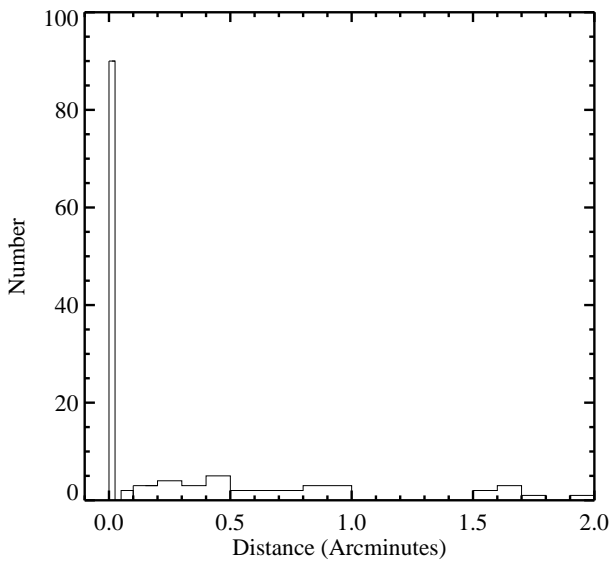}
\caption{Histogram of distances between objects in the SL cluster
  sample and the matching spectra in the spectral database.  There are
  several objects that have poor matches.  Objects with match distances
  above 1.5 arcseconds and below 2 arcminutes were manually inspected.  
  The smallest bin contains distances less than 1.5 arcseconds.  There are 90 objects
  in this bin but 6 of these were removed after a visual inspection deemed them 
  not clearly real lenses, yielding the number 84 cited in the text.}
\label{hist}
\end{figure}

\subsection{Matching SL Cluster Coordinates to BCG Spectra}
In order to obtain the spectra for the BCGs of interest from the SDSS 
data set, the SL sample cluster coordinates were matched with 
the MPA-JHU release of spectrum measurements from SDSS DR7.  The SL
sample coordinates come from a visual inspection of the field where the centers of 
mass of the clusters are approximated by eye.  When an arc forms
around an obvious BCG, the centroid of the BCG is assigned as the center of mass.  
However, in cases where there is no obvious BCG, the
centroid of the arc itself is used.  As a result, we first look for
exact matches to BCG spectra and then manually inspect the near-match
cases.  Coordinates of the SL clusters were compared to the positions in the spectral  
catalog to find the separation between each sample object and all the
objects in the spectral database.  The match for each sample cluster is then the object 
in the spectral database that is the smallest distance away from that
SL cluster.  84 SL clusters had lensing centers that matched those of spectra 
in the database to within 1.5 arcseconds, where the 1.5 arcsecond cut is 
motivated by the size of the SDSS spectroscopic fiber aperture (3 arcsecond 
diameter).

\begin{figure}[htp]
\epsscale{1.05}
\plotone{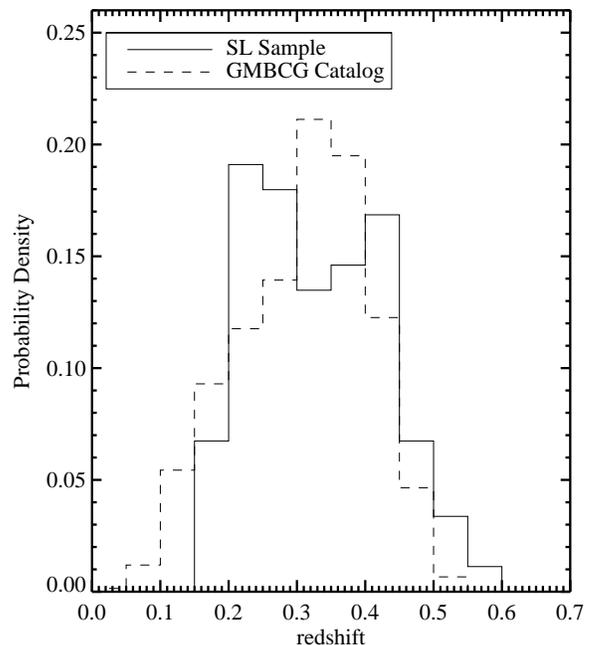}
\caption{Redshift distribution of the clusters in the SL sample that
  have SDSS spectra and are used in the analysis (solid) compared with
  the redshift distribution of the GMBCG catalog (dashed).}  
\label{zdist}
\end{figure}

Figure \ref{hist} shows the
histogram of distances for the matching process between the SL cluster 
sample and the spectroscopic database.  It is clear
that some SL clusters do not have matches with the spectra file. 
Images of those non-matches for which the match distance is greater than 1.5
arcseconds but less than 2 arcminutes were manually inspected to
determine if there are any appropriate bright cluster member galaxies with spectra.  For
example, some clusters may have multiple bright galaxies in the core,
all located close to the center of mass of the cluster.  The galaxy
corresponding to the SL sample coordinates might not have a spectrum
but another galaxy nearby, that is also part of the central mass
distribution, might have one.  As mentioned above, some of
the SL coordinates are actually centroids of giant arcs so the
corresponding BCG with a spectrum must be manually determined.  5 of the
moderately matching systems were included in the final sample. 
The final sample thus results in 89 clusters.  The
spectroscopic redshift distribution of these clusters is shown in
Figure \ref{zdist}.

\subsection{Optically Selected Galaxy Cluster Catalog}
To compare our results to the total cluster population, we are using
the GMBCG catalog \citep{Hao2010} which was also used by \citet{McDonald2011} who
studied the evolution of optical line emission in the total population. 
The GMBCG catalog was created by searching for BCGs
and the red sequence to
find galaxy clusters from SDSS DR7 producing a catalog of over 55,000
galaxy clusters in the redshift range 0.1 $< z <$ 0.55.  The
spectroscopic redshift
distribution of all the GMBCG clusters with SDSS spectra is shown in
Figure \ref{zdist}.  We can also compare the range in cluster masses spanned by the SL and GMBCG samples using a galaxy cluster richness estimator.  Most of the SL clusters in our sample have measured richnesses from the GMBCG catalog.  For the remaining clusters we used a similar procedure to that used by \citet{Hao2010} to measure richness so that we could compare the richness distribution between the two samples.  We note that the richness distribution of our SL sample represents a subset of the richness distribution of the GMBCG catalog weighted towards higher richness.  The mean richness and $1-\sigma$ uncertainties of the SL sample is $22^{+48}_{-11}$ and the mean and uncertainty of the GMBCG is $12^{+6}_{-11}$.  The richnesses of the SL sample span the range $2-87$ and the GMBCG richnesses range from $8$ to $143$ with only $0.1\%$ greater than 87.  The SL sample is drawn from the full range of GMBCG richnesses with a preference for higher richness as expected from simulations \citep{Hennawi2007, Meneghetti2010}.      

\subsection{SDSS Data}
The relevent data include spectroscopic 
redshifts, emission line flux measurements and $4000$\AA\ break ratios.  
The lines of interest in this work are H$\beta$, [OII]$\lambda\lambda$3727, and
[OIII]$\lambda$$\lambda$5007.  We do not use H$\alpha$ because at $z
\backsim 0.4$ it is redshifted out of the wavelength coverage of
SDSS.  The [OII] line is a good tracer of star formation
rates over the redshift range of our sample \citep{Kennicutt1998,
  Kewley2003} and stays within the wavelength coverage of SDSS which is
3800\AA\ - 9200\AA\footnote{http://www.sdss.org/dr7/instruments/spectrographs/index.html}.  
Also, in our redshift range the [OII] line stays
blueward of bright sky lines that exist redward of 7300\AA, which can cause
residuals from sky subtraction.  

An important property of the SDSS spectroscopic database is that the 
3$^{\prime\prime}$ spectroscopic fiber aperture encompasses different physical regions on 
the sky at different redshifts, as the angular diameter distance changes with 
redshift. For nearby BCGs, for example, the fiber aperture will only encompass 
a fraction of the total area of the BCG, and would therefore fail to detect any line 
emission from extended, often filamentary, regions beyond the physical radius 
probed by the SDSS fiber.  \citet{McDonald2011}
showed that above a redshift of about 0.3, the fiber aperture
encompasses nearly the total H$\alpha$ emission from a sample of
galaxy clusters.  But for $z < 0.3$ it is essential that an aperture
correction be performed.  Two aperture corrections were used in this
work.  \citet{McDonald2011} derived a universal L$_{H\alpha}(r)$
profile based on low-z, well-resolved systems to determine the
fraction of emission outside the aperture.
The second correction assumes that the mean H$\beta$ luminosity should
be constant with distance.  Thus, the only change in H$\beta$
luminosity should be due to the aperture encompassing different
physical diameters. As in \citet{McDonald2011}, we find that the two different 
aperture corrections produced consistent results.

%\vspace{1cm}
\section{Cooling Signatures in Strong Lensing Galaxy Clusters}
\subsection{Evolution of Emission in the Range $0.1 < z < 0.6$}
To understand the evolution of [OII] line emission in strong lensing
galaxy clusters we must determine the fraction of SL galaxy clusters
that exhibit [OII] line emission as a function of some redshift bin.
To do this, for each SL galaxy cluster we calculate the probability,
assuming Gaussian statistics,
that the line luminosity is above a certain threshold.  Following
\citet{McDonald2011}, the condition for strong [OII] emission is $L_{[OII]} >
3.1 \times 10^{40}$ erg s$^{-1}$ and the condition for weak [OII] emission is
$7.8 \times 10^{39}$ erg s$^{-1} < L_{[OII]} < 3.1 \times 10^{40}$ erg
s$^{-1}$.  In a given
redshift bin, the fraction of SL galaxy clusters with weak or strong [OII]
emission is given by the average of the individual
probabilities for each cluster in that bin.  To avoid confusing
optical line
emission from warm gas in BCGs with AGN activity, if [OIII]/H$\beta >
3$ (i.e. Seyfert galaxy where [OII] emission is not necessarily from star
formation) for a particular cluster, the cluster is classified as non-emitting
and the probability of being an [OII] emitter is set to zero.  14 of
the SL clusters in the sample fall into this category of non-emitting.

\begin{figure}[htp]
\epsscale{1.1}
\plotone{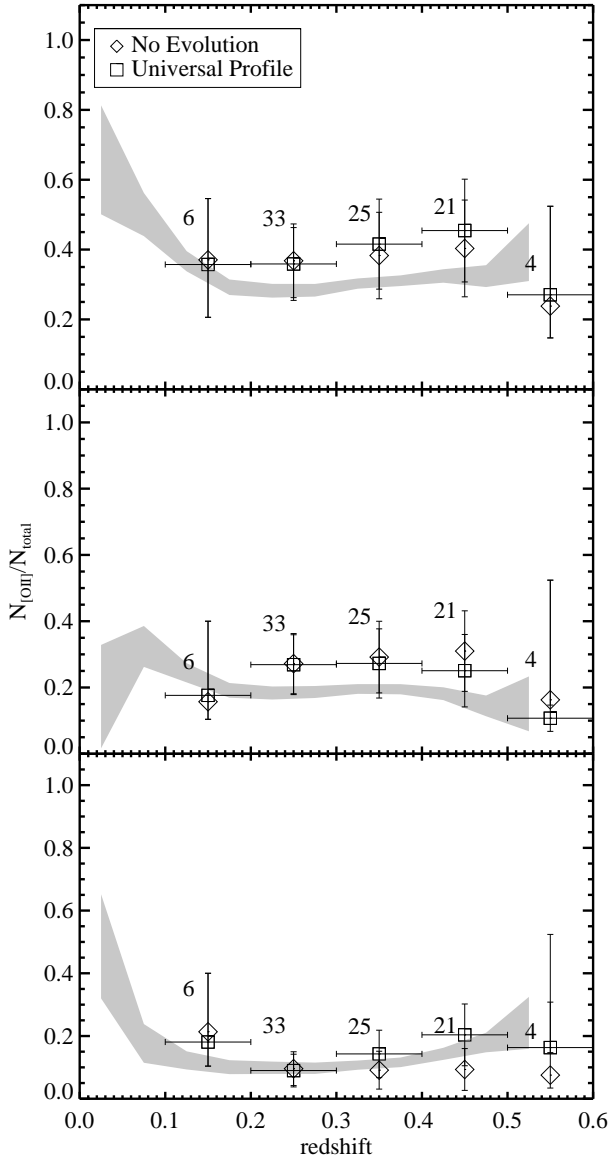}
\caption{The fraction of SL clusters with all (top),
  weak (middle), and strong (bottom) [OII] emission.  The two aperture 
  corrections mentioned in the
  text have been applied and are here referred to as ``Universal
  Profile'' and ``No Evolution''.  These two corrections agree well.  
  The errors for the three middle bins are the standard
  deviations of the means in each bin calculated using Poisson
  statistics.  The lowest and highest bin errors were calculated using
  binomial methods outlined by \citet{Cameron2011}.  The numbers near
  the data points indicate how many
  clusters are in each bin.  The gray areas here correspond to [OII]
  emission evolution for the ``no evolution'' aperture correction 
  applied to the GMBCG sample in \citet{McDonald2011}.}
\label{OLEplot}
\end{figure}

Figure \ref{OLEplot} shows the
fraction of SL galaxy clusters with all, weak, and strong [OII]
emission in the central galaxy.  This is the evolution of [OII] emission for
89 SL galaxy clusters in the range $0.1 < z < 0.6$.  The over-plotted
gray areas represent the evolution of emission for the GMBCG catalog from
\citet{McDonald2011}. The statistical agreement between the SL sample and
the GMBCG catalog indicates that the fraction of central galaxies in
SL clusters with bright [OII] emission as a function of redshift
differs little from the general cluster population.  The trend
of a constant fraction of optical line emission for $z > 0.2 $
in the general cluster population appears to be mirrored in the strong
lensing cluster sample.  If a large fraction of SL galaxy clusters
showed strong [OII] emission, then this would suggest that baryonic
cooling plays an important role in increasing SL cross-sections. 
Instead, we find no evidence for an enhancement in [OII] emission and
thus, baryonic cooling, in strong lensing selected clusters. The mean [OII] 
fractions (for each of the all, weak, and strong cases) that we compute for 
both the SL sample and the GMBCG catalog at $z > 0.2$ are given in 
Table \ref{SL}.

\begin{table}[htp]
\caption{Mean all, weak, and strong [OII] emission fractions for $z >
  0.2$ using the universal profile (Univ) and no evolution (NoEv) aperture corrections.  
  Errors are 1-$\sigma$.}
\begin{center}
\begin{tabular}{l|l|l|l|l}

   & \multicolumn{2}{c}{SL Sample} &  \multicolumn{2}{c}{GMBCG
    Catalog} \\
\hline
   & Univ & NoEv & Univ & NoEv \\  
\hline
weak & .26$\pm$.06 & .28$\pm$.06 & .183$\pm$.003 & .189$\pm$.003 \\
\hline
strong & .14$\pm$.04 & .09$\pm$.03 & .125$\pm$.003 & .110$\pm$.002 \\
\hline
all & .40$\pm$.07 & .38$\pm$.07 & .308$\pm$.004 & .299$\pm$.004 \\
\label{SL}
\end{tabular}
\end{center}
\end{table}

%\vspace{1cm}
\subsection{Probing Star Formation Using the 4000\AA\ Break Ratio}
\begin{figure}[tp]
\epsscale{1.05}
\plotone{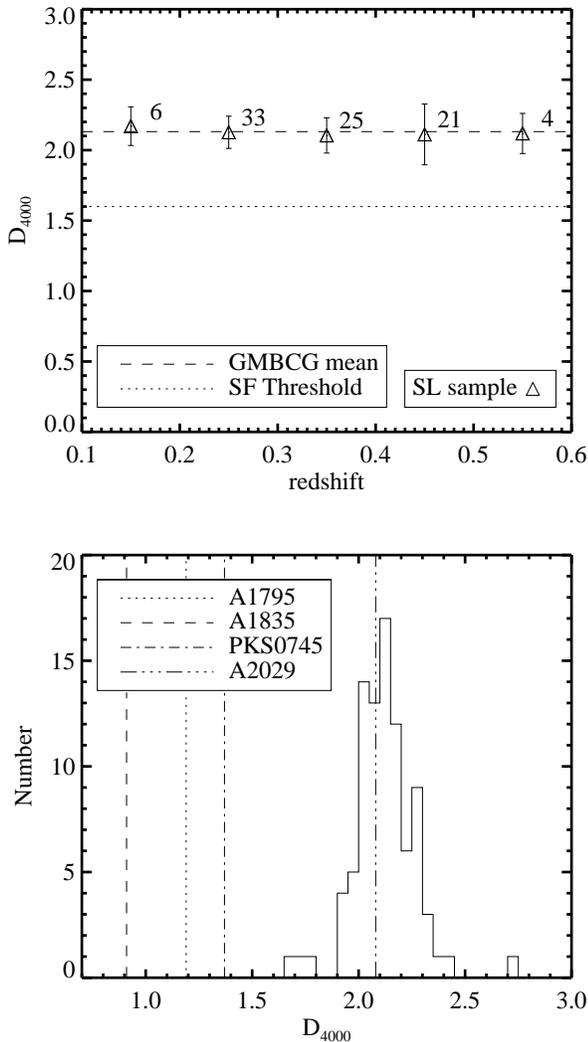}
\caption{The top plot shows the evolution of the 4000\AA\ break ratio
  (D$_{4000}$) in our
  sample plotted with the mean ratio for the GMBCG catalog (dashed
  line).  The dotted line is the median of the D$_{4000}$ distribution from
  \citet{Kauffmann2003}, indicating the approximate threshold between
  star forming and non-star forming galaxies.  The evolution in the SL
  sample does not deviate from the typical ratio of the general
  cluster population.  Errors are calculated from Poisson counting
  statistics.  The bottom plot shows the distribution of the break
  ratio in the SL sample. PKS0745, A1795, and A1835 are classical strong cool
  cores and A2029 is a non-cool core. Values for PKS0745, A1795, and
  A2029 come from \citet{Johnstone1987} and the value for A1835 is
  from SDSS.}
\label{d4000}
\end{figure}

As a check on our results we can investigate the specific star
formation rate (sSFR), another tracer of ongoing cooling, in each strong
lensing BCG and compare it to the rate in the total population.  This sSFR
must be independent from our flux measurements to contain new
information so we use the 4000\AA\ break index provided by the
MPA-JHU data release as a tracer for specific star formation rate.  The
4000\AA\ break index is the ratio of the mean flux in the range
4000\AA\ - 4100\AA\ to the mean flux in the range 3850\AA\ - 3950\AA\
\citep{Brinchmann2004}.  Objects with low star
formation, and thus few young, blue stars, will have strong 4000\AA\ break
ratios.  In Figure \ref{d4000} we plot the mean 4000\AA\ break of our
SL clusters in five redshift bins as well as the mean 4000\AA\ break
of the GMBCG catalog.  There is no deviation from the GMBCG catalog,
indicating that SL clusters exhibit the same specific rate of star formation as
the general population of BCGs.  This is consistent with our results above
that found that [OII] line emission in SL clusters deviates little
from the total population.  

To understand the break strength distribution of the SL sample we also
plot a histogram of the distribution in Figure \ref{d4000}.  The lack
of strong bimodality suggests that the SL sample clusters are not
forming many stars in their cores.  Typical star forming galaxies
tend to have break
strengths of $\backsim$ 1.3 \citep{Kauffmann2003}.  The division
between star forming and non-star forming galaxies occurs around a
break strenth of 1.6 \citep{Kauffmann2003}.  The SL cluster BCGs
have break strength values indicating they are predominantly non-star
forming.  The vertical lines in the histogram of Figure \ref{d4000}
indicate the 4000\AA\ break strength values for various classical cool cores
and non-cool cores.  PKS0745, A1795, and A1835 are strong cool cores
whereas A2029 is a non-cool core.  The strong cool cores tend to have
values well below the SL sample while A2029 has a value $\backsim$ 0.1
away from the SL sample mean indicating that SL sample clusters are
not exhibiting the typical break strength values of cool core clusters.        

\subsection{[OII] Emission Fraction As a Function of Strong Lensing Cross-Section}
To better characterize the above results, we investigate whether or not
clusters with a larger strong lensing cross-section show stronger
emission. We use an observationally defined quantity, R$_{arc}$, for 
each cluster lens as a proxy for strong lensing cross-section.  We define R$_{arc}$ 
as the radial separation between the arcs and the center of mass of the SL clusters.  R$_{arc}$ is 
an observable that is simple to measure for our entire 
sample, and which provides an approximate estimate of the Einstein radius. 
The Einstein radius describes the critical curve for a given 
strong lens, and is defined analytically as the location in the lens plane where 
the formal magnification of a source distorted by a lens goes to infinity 
\citep{Schneider1992}. In the simplest 
case of a spherically symmetric lensing potential and perfect alignment 
between the source, lens, and observer, the source is re-imaged into a ring 
described by the Einstein radius. The radius of this ring is the Einstein radius,
$\theta_{E}$, and is given by:

\begin{equation}
\theta_{E} = \sqrt{\frac{4GM}{c^2}\frac{D_{LS}}{D_{L}D_{S}}}
\end{equation}

where $G$ is Newton's gravitational constant, $M$ is the mass of the
lensing cluster, $c$ is the speed of light, D$_{LS}$ is the distance
between the lens and the source, D$_{L}$ is the distance between the
observer and the lens, and D$_{S}$ is the distance between the
observer and the source.

Physically realistic lensing systems have 
critical curves with much more complex morphologies, but the Einstein 
radius for such systems can still be defined and measured as the radius 
of a circle which has the same area on the sky as the area contained 
within the critical curve. The size of the critical curve provides a 
measurement of the ``strong-ness'' of a strong lens, where the SL cluster 
population consists of a broad range of structures ranging from the rarest 
super-lenses with extreme strong lensing cross-sections, to the more 
numerous marginal strong lenses. 

Detailed strong lensing reconstructions of the critical curves for 
our cluster lens sample is observationally unfeasible as it would require 
extensive follow-up observations. However, rather than model the critical 
curve for each SL cluster, it is also possible to define a simple observable 
quantity by fitting an ellipse to a multiply imaged source -- or giant arc -- 
and measure the radius corresponding to a circle with an area equal to 
the area of the fitted ellipse (R$_{arc}$). Tests in simulations show that 
this quantity has a large intrinsic uncertainty when used to estimate 
the Einstein radius for an individual lens system, but that on average it 
correlates with Einstein radius \citep{PuchweinHilbert2009}. 
We can therefore use R$_{arc}$ for our SL cluster sample to sort lenses 
approximately by the size of their strong lensing cross-section. This sorting 
allows us to probe whether baryonic cooling processes may be helping to 
drive up strong lensing cross-sections within a subset of the total cluster 
lens population.

We estimate the radial separation, R$_{arc}$, of the arcs from the
center of mass of the cluster in each SL cluster from optical follow-up images 
taken with the Mosaic Camera (MOSCA) on the 2.5m Nordic Optical
Telescope.  In each image the center of mass of the cluster (usually the BCG) as well as the
arcs are located. The fitting program mpfitellipse \citep{More1978, Markwardt2009}
is then used to fit an ellipse to the curvature of the arcs to recover a rough 
estimate of the critical curve for each cluster lens.  For the measured radial 
separations to be useful as a way to sort and compare members of our 
sample, they must be scaled to remove the distance dependence of each 
measurement.  This is accomplished by scaling each measurement by:

\begin{equation}
N = \sqrt{\frac{\frac{D_{L_{0}S_{0}}}{D_{L_{0}}D_{S_{0}}}}{\frac{D_{LS}}{D_{L}D_{S}}}}
\end{equation} 

where D$_{LS}$, D$_{L}$, D$_{S}$ are the relevant distance values for each
particular cluster, and D$_{L_{0}S_{0}}$, D$_{L_{0}}$, D$_{S_{0}}$ are
values for a fiducial lens configuration. Because the source redshifts for many 
of our individual SL systems are unknown, we use the typical source 
redshift as measured in the literature, z$_{s} = 2 \pm 1$ 
\citep{Bayliss2011,Bayliss2012}.  The source redshift uncertainty for each 
individual lens system produces a systematic uncertainty in the final scaled 
R$_{arc}$ values for our SL cluster sample, but this uncertainty is quite small 
($\sim +3\% - 8\%$ for a lens redshift of 0.3, the median of our sample) and does not impact our results. 

\begin{figure}[btp]
\epsscale{1.1}
\plotone{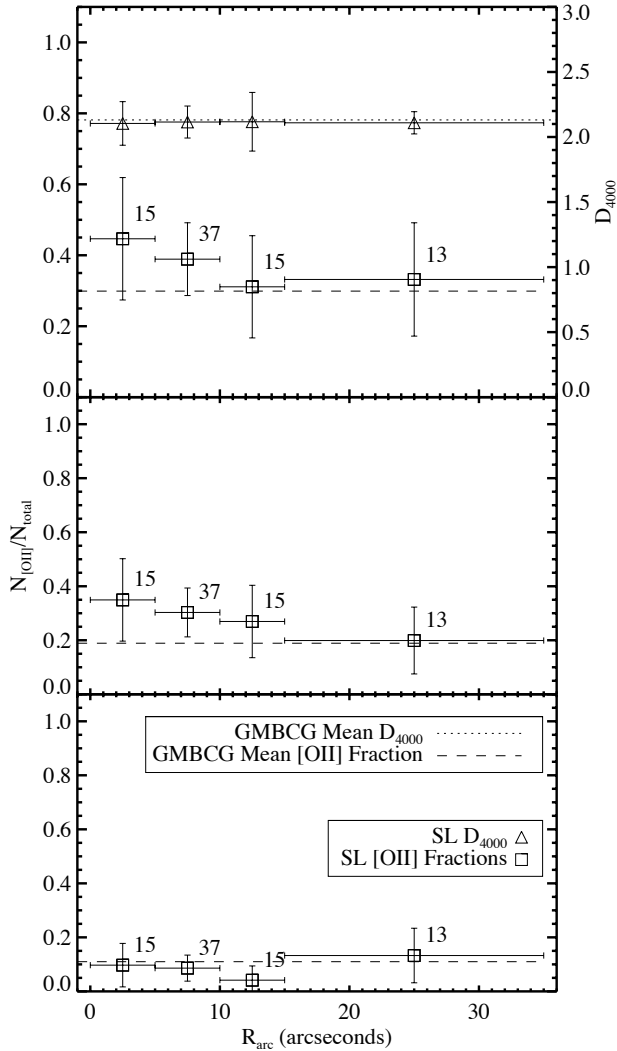}
\caption{This plot shows the fraction of SL clusters with all (top),
  weak (middle), and strong (bottom) [OII] emission in four bins of
  R$_{arc}$.  Only
  those SL clusters with $z > 0.2$ are considered here.  The
  dashed line represents the mean GMBCG [OII] emission fraction.  All
  [OII] fractions in this plot were calculated with fluxes corrected
  using the ``no evolution'' aperture correction.  The top plot also
  shows the 4000\AA\ break ratios (D$_{4000}$) as a function of
  R$_{arc}$.  The dotted line represents the mean GMBCG break ratio
  for $z > 0.2$.}
\label{einstein}
\end{figure}

With the scaled R$_{arc}$ measurements, one can then determine
the [OII] emission fraction as a function of R$_{arc}$ ($\backsim
\theta_{E}$).
Figure \ref{einstein} shows the fraction of SL clusters with all,
weak, and strong [OII] emission in four bins of R$_{arc}$.  In this
plot only those SL clusters with $z > 0.2$ were included because this
is where there is no evidence of changing [OII] emission fractions.
From Figure \ref{einstein} it seems that there is no statistically significant
dependence of [OII] emission on Einstein radii above about 10
arcseconds.  Below this value there is a slight increase in the
fraction of weak [OII] emitters whereas the strong [OII] fraction is
consistent with no dependence.  For weak emission the data points in
the bins below 10 arcseconds deviate from the GMBCG mean by about 1-$\sigma$
and for total emission the data points deviate from the GMBCG mean by less than
1-$\sigma$ and are thus not statistically robust deviations. 
Figure \ref{einstein} also
shows the 4000\AA\ break strengths as a function of R$_{arc}$, which
show no deviation from the GMBCG mean break strength and no 
evidence for variation in the break strength as a function of R$_{arc}$.
Clusters with large Einstein radii exhibit optical tracers of baryonic cooling 
in their cores with the same frequency as clusters with small Einstein radii, 
and also as the total cluster population.

\section{Discussion}
Figure \ref{OLEplot} demonstrates that the fraction of strong lensing galaxy clusters over the range $0.2 < z <
0.6$ with [OII] line-emitting BCGs is constant and shows no statistically significant deviation from the total cluster population, suggesting that baryonic cooling is not enhanced in SL clusters over the general cluster population. 
Figure \ref{d4000}
supports this conclusion by showing that there is no evolution in 4000\AA\
break ratios and that they match the mean ratio in an optically selected 
sample of galaxy clusters -- the GMBCG catalog.
Furthermore, the typical D$_{4000}$ value for the SL sample is consistent 
with non-cool cores that are not forming many stars in the BCG.  If ongoing cooling 
were playing a continuing role in generating efficient SL clusters then 
we would expect to see some evidence of enhanced cooling in the form of intermediate 
temperature (10$^{4}$K) gas or ongoing star formation 
\citep[e.g.,][]{Edge2001, O'Dea2008, McDonald2010, McDonald2011a, McDonald2011b, Tremblay2012}, as 
traced by optical emission or the 4000\AA\ break in the cluster cores. 

We find no evidence for such an enhancement; instead, our analysis suggests that cool cores are no more prevalent in strong lensing clusters than in the general cluster population.  Our results argue 
that baryonic cooling associated with cool core activity is not an efficient mechanism for dramatically increasing strong lensing cross-sections in galaxy clusters.  
\citet{Rozo2008} and \citet{Mead2010}
found that simulations which include baryonic cooling can increase
strong lensing cross-sections of simulated galaxy clusters by factors of $\backsim$ 2-3.  
These scenarios require a ``runaway" cooling flow which causes dark matter to condense 
in the core by sufficient amounts to alter the total matter density profile and the strong lensing 
properties of the cluster.  Since runaway cooling flows are not observed, it is evident that 
other factors, like AGN feedback, act on sufficiently short timescales to prevent runaway 
cooling and unrealistically cuspy gas density profiles \citep{Best2005, McNamara2007, 
Fabian2012, McNamara2012}.  Otherwise, we would observe the effects of this runaway 
cooling in the form of massive starbursts.  This feedback scenario is consistent with recent 
studies \citep[e.g.][]{Mead2010, Killedar2012} that found that simulations which
include models of AGN feedback, together with cold dark matter and
gas dynamics, show less significant increases in strong lensing 
cross-sections.  This agreement between observational and simulation-based results is 
encouraging, and suggests that the current generation of cosmological simulations 
include feedback models that are sufficiently sophisticated to recover the impact of 
baryonic processes on the total matter distribution in cluster cores.

Our results are also interesting in the context of recent work in which the slopes of the central 
density profiles in a small sample of relaxed clusters were estimated from multi-wavelength 
observations \citep{Newman2013}. The selection of the clusters studied 
by \citet{Newman2013} complicates a direct comparison between their conclusions and the 
results of our work, which uses a large generic strong lensing selection. 
\citet{Newman2013} found that the observed density profiles of their seven clusters are in 
good agreement with the predictions from dark matter (DM) only simulations, measuring 
total density profiles in the cores of seven clusters with slopes that match cold dark matter 
(CDM) simulations. They argue that dynamical heating is a possible mechanism for offsetting 
any effects that baryonic contraction might have on the matter distribution in massive cluster 
cores.

It makes sense that the results of such a mechanism would be observable in a 
sample of clusters that was chosen specifically to be dynamically relaxed and undisturbed, 
where the total matter distribution in the cores (baryonic$+$DM) has had the opportunity to 
virialize. However, the strong lensing selection of the SGAS cluster lens sample does not 
preferentially select for relaxed systems, and in fact there is evidence suggesting that 
dynamically disturbed and merging systems should be well-represented in a strong lensing 
selected cluster sample \citep{Torri2004,Oguri2013}. The matter distribution in the cores of 
such a sample should not necessarily be expected to have the same average profile 
properties as a sample that is selected to be relaxed.

Having noted the different selection criteria for our sample and that of \citet{Newman2013}, 
we do note that there is broad agreement between our results and those of 
\citet{Newman2013} in that neither result favors a scenario in which baryonic cooling is 
acting to steepen the matter distributions in the cores of clusters. It therefore 
follows that it is not reasonable to invoke baryonic cooling as a dominant explanation for 
the apparent discrepancies between observed and predicated arc abundances 
\citep{Bartelmann1998,luppino1999,ZaritskyGonzalez2003,Gladders2003,li2006}.

We note that in Figure \ref{einstein} there is a marginal increase (at the $\sim 1-\sigma$ level) 
in the fraction of strong lensing clusters with R$_{arc} < 10^{\prime\prime}$ exhibiting weak 
[OII] emission. The observable R$_{arc}$ correlates strongly with Einstein Radius, 
which itself correlates with the total mass of the cluster lens, so that the 
R$_{arc} < 10^{\prime\prime}$ bin will include, on average, the lower-mass cluster lenses 
in our sample. This marginal increase is in qualitative agreement with the suggestion that 
baryonic cooling could be responsible for small excesses in the concentration parameters 
measured for lower-mass and smaller Einstein radius strong lensing selected clusters by 
\citet{Oguri2012}. However, neither the increase in optical line emission that we measure, nor the 
excess concentrations in \citet{Oguri2012} are statistically robust (i.e. $>$ 2-$\sigma$), 
and we refrain from claiming that the combination of these two results can be interpreted as 
strong evidence for cooling baryons driving up concentrations in low-mass or small 
Einstein radius strong lensing clusters. These marginal excesses in optical line emission and concentration 
could, however, reflect consistency with the expectation from simulations that gas cooling 
may more strongly affect clusters with lower masses where the cooling mass in the core can 
comprise a larger fraction of the total mass \citep{Rozo2008,Killedar2012}.

\section{Summary}
In this work, we searched for optical line emission and recent star
formation in a sample of 89
strong lensing galaxy clusters to probe whether or not baryonic
cooling processes significantly affect the mass density profiles of
clusters.  Using published SDSS spectral data for the BCGs of the
SL clusters we have calculated the fraction of SL clusters with
[OII] line emission as a function of redshift.  We find that the
evolution of [OII] line emission in the SL sample is constant for
$z > 0.2$ and that there is no statistically
significant difference
between the SL sample and the general cluster population.  The
4000\AA\ break ratio in the SL sample also matches the general
population, indicating that the average specific star formation rate is similar
between the two populations.  We also sorted the SL cluster sample 
by R$_{arc}$ -- an observable that correlates strongly with Einstein 
radius -- to look for trends in the optical tracers of gas cooling as a 
function of the individual lens cross-sections. We find that [OII] 
line emission fractions and  4000\AA\ break ratios showed no 
significant dependence on Einstein radius, suggesting that baryonic 
cooling does not play a large role increasing strong lensing 
cross-sections among either the small or large strong lensing 
cross-section end of the total cluster lens population. The results of this 
work combined with the well-studied correlations between ICM 
cooling and BCG star formation and line emission argue strongly 
that baryonic cooling associated with cool core activity does not significantly influence the dark matter 
distribution to steepen the mass density profile in the cores of strong 
lensing galaxy clusters.

\acknowledgments
This work is supported in part by the National Science Foundation
Research Experiences for Undergraduates (REU) and Department of Defense
Awards to Stimulate and Support Undergraduate Research Experiences
(ASSURE) programs under grant number 0754568 and by the Smithsonian
Institution.  Part of this work was based on observations 
made with the Nordic Optical Telescope, operated
on the island of La Palma jointly by Denmark, Finland, Iceland,
Norway, and Sweden, in the Spanish Observatorio del Roque de los
Muchachos of the Instituto de Astrofisica de Canarias.  M. B. B. acknowledges support from the NSF Astronomy Division under grant number AST-1009012.  M. M. acknowledges support provided by NASA through a Hubble Fellowship grant from STScI.  M.D.G. thanks the Research Corporation for support of this work through a Cottrell Scholars award.  We would like to thank 
Jonathan McDowell and Marie Machacek for helpful feedback on early drafts of this paper.

\bibliographystyle{apj}

\end{document}